\documentclass[12pt]{article}
\usepackage{amsfonts}
\usepackage{amsmath}
\usepackage{amssymb}
\usepackage{amsthm}
\usepackage{graphicx}
\usepackage{latexsym}
\usepackage{xy}
\usepackage{verbatim}
\usepackage{bm}
\usepackage{hyperref}

\parskip=\baselineskip
\begin{document}

\begin{titlepage}
\begin{flushright}
\end{flushright}
\vskip 1.5in
\begin{center}
{\bf\Large{CHIRAL GRAVITY IN THREE DIMENSIONS}} \vskip 0.5in {Wei
Li, Wei Song\footnote{Permanent address: \textit{Institute of
Theoretical Physics Academia Sinica, Beijing 100080, China.}} and
Andrew Strominger}
\vskip 0.3in
{\small{\textit{ Center for the Fundamental Laws of Nature\\
Jefferson Physical Laboratory, Harvard University, Cambridge MA
02138, USA}}}

\end{center}
\vskip 0.5in

\baselineskip 16pt
\date{}

\begin{abstract}
Three dimensional Einstein gravity with negative cosmological
constant $-1/\ell^{2}$ deformed by a gravitational Chern-Simons
action with coefficient $1/\mu$ is studied in an asymptotically
$AdS_3$ spacetime. It is argued to violate unitary or positivity for
generic $\mu$ due to negative-energy massive gravitons. However at
the critical value $\mu\ell=1$, the massive gravitons disappear and
BTZ black holes all have mass and angular momentum related by  $\ell
M=J$. The corresponding  chiral quantum theory of gravity is
conjectured to exist and be dual to a purely right-moving boundary
CFT with central charges $(c_L,c_R)=(0,3\ell /G)$.

\end{abstract}
\end{titlepage}
\vfill\eject

\tableofcontents
\section{Introduction}
In three dimensions, the Riemann and Ricci tensor both have the same
number (six) of independent components.
 Hence Einstein's equation, with or without a cosmological constant $\Lambda$, completely constrains the geometry
 and there are no local propagating degrees of freedom. At first sight this makes the theory sound
 too trivial to be interesting. However in the case of a negative cosmological constant, there are asymptotically
  $AdS_3$ black hole solutions \cite{Banados:1992wn} as well as massless gravitons which can be viewed as
  propagating on the boundary. These black holes obey the laws of black hole thermodynamics and have an
 entropy given by one-quarter the horizon area. This raises the interesting question: what is the microscopic
 origin of the black hole entropy in these ``trivial" theories?

In order to address this question one must quantize the theory. One
proposal \cite{Witten:1988hc} is to recast it as an
$SL(2,\mathbb{R})_L\times SL(2,\mathbb{R})_R$ Chern-Simons gauge
theory with $k_L=k_R$. Despite some effort this approach has not
given a clear accounting of the black hole entropy (see however
\cite{Fjelstad:2001je} for an interesting attempt).
    So the situation remains unsatisfactory.

 More might be learned by deforming the theory with the addition of the gravitational  Chern-Simons term
 with coefficient ${1 \over \mu}$ \cite{Deser:1981wh}\cite{Deser:1982vy}. The resulting theory is known as the topologically
 massive gravity (TMG) and contains a local,
 massive propagating degree of freedom, as well as black holes and massless boundary gravitons. The addition of the Chern-Simons
  term leads to more degrees of freedom because it contains three, rather than just two,  derivatives of the metric.
  It is the purpose of this note
  to study this theory for negative $\Lambda=-1/\ell^2$.
 We will argue that the theory is unstable/inconsistent for generic $\mu$: either the massive gravitons or BTZ black holes have
 negative energy.  The exception occurs when the parameters obey  $\mu\ell=1$, at which point several interesting phenomena
 simultaneously arise:

 (i) The central charges of the dual boundary CFT become $c_L=0$, $c_R=3\ell/G$.

  (ii) The conformal weights as well as the wave function of the massive
  graviton, generically $\frac{1}{2}(3+\mu\ell,-1+\mu\ell)$,
  degenerate with those of the left-moving weight $(2,0)$ massless boundary graviton.
  They are both pure gauge, but the gauge transformation parameter does not  vanish at infinity.

  (iii) BTZ black holes and all gravitons have non-negative masses. Further the angular momentum is fixed in terms of the mass to be
  $J=M\ell$.

  This suggests the possibility of a stable, consistent theory at $\mu\ell=1$ which is dual to a holomorphic boundary CFT (i.e. containing only
  right-moving degrees of freedom) with $c_R=3\ell/G$. The hope - which remains to be investigated - is that for a suitable choice of
   boundary conditions the zero-energy left-moving excitations can be discarded as pure gauge. We will refer to this theory as 3D chiral gravity.
As we will review herein, if such a dual CFT exists, and is unitary,
an application of the Cardy formula gives a microscopic derivation
of the black hole entropy
\cite{Strominger:1997eq}\cite{krauslarsen}\cite{Kraus:2005zm}.

 Related recent work \cite{Witten:2007kt}-\cite{Maloney:2007ud} has considered an alternative deformation of pure 3D gravity, locally described by the
 $SL(2,\mathbb{R})_L\times SL(2,\mathbb{R})_R$ Chern-Simons gauge theory with $k_L\neq k_R$.
   This is a purely topological theory with no local degrees
 of freedom and is not equivalent to TMG. It contains all the subset of solutions of TMG which are Einstien metrics but not the massive gravitons.
 It is nevertheless possible that the arguments given in  \cite{Witten:2007kt} (adapted to the case
 $k_L=0$ as in \cite{Manschot:2007zb}) which are quite general apply to the chiral gravity discussed herein. Indeed discrepancies with the semiclassical analysis mentioned in
 \cite{Witten:2007kt}-\cite{Maloney:2007ud}  disappear for
 the special case $k_L=0$.\footnote{Although we do not study the Euclidean theory herein the relation $M\ell=J$ for Lorentzian BTZ black holes in chiral gravity suggests that the saddle point action
 will be holomorphic.}
Moreover the main assumption of \cite{Witten:2007kt} - holomorphic factorization of the partition function - is simply a consistency requirement
 for chiral gravity because there are only right movers.

 The paper is organized as follows. Section 2 gives a brief review of
 the cosmological TMG and its $AdS_3$ vacuum solution, and shows that the theory is purely chiral at the special value of $\mu=1/\ell$. Section 3 describes the linearized gravitational excitations around
  $AdS_3$. Section 4 shows how the positivity of energy imposes a stringent constraint on the allowed value of $\mu$. We end with a short summary and discussion of future directions.

\section{Topologically Massive Gravity}

\subsection{Action}
The action for TMG (topologically massive gravity) with a negative
cosmological constant is \cite{DeserTekin1}
\begin{equation}
I= \frac{1}{16\pi G}\int d^3x \sqrt{-g}
(R-2\Lambda)+\frac{1}{16\pi G\mu}I_{CS}
\end{equation}
where
 $I_{cs}$ is the Chern-Simons term
\begin{equation}
I_{cs}=\frac{1}{2}\int
d^3x\sqrt{-g}\epsilon^{\lambda\mu\nu}\Gamma^\rho_{\lambda\sigma}
[\partial_\mu\Gamma^\sigma_{\rho\nu}+\frac{2}{3}\Gamma^\sigma_{\mu\tau}\Gamma^\tau_{\nu\rho}]
\end{equation}
and we take $\Lambda$ negative.
We have chosen the sign in front of the Einstein-Hilbert  action so that BTZ black holes have positive energy for large $\mu$,
 while the massive gravitons will turn out to have negative energy for this choice.\footnote{This contrasts with most of the literature which chooses the opposite sign in order that massive gravitons have
 positive energy (for large $\mu$).}
 The equation of motion is
\begin{equation}\label{eom}{\cal G}_{\mu\nu}+\frac{1}{\mu}C_{\mu\nu}=0\
,\end{equation} where $\cal{G}_{\mu\nu}$ is the cosmological-constant-modified
Einstein tensor:
\begin{equation}
{\cal G}_{\mu\nu}\equiv R_{\mu\nu}-\frac{1}{2}g_{\mu\nu}R+\Lambda
g_{\mu\nu}\
\end{equation}
and $C_{\mu\nu}$ is the Cotton tensor
\begin{equation}
C_{\mu\nu}\equiv \epsilon_\mu\,^{\alpha\beta}\nabla_\alpha
(R_{\beta\nu}-{1\over4} g_{\beta\nu} R).
\end{equation}
Einstein metrics with $\cal{G}_{\mu\nu}=~$0 are a subset of the general solutions of (\ref{eom}).

\subsection{$AdS_3$ vacuum solution }

TMG has an
$AdS_3$ solution:
\begin{equation}\label{AdS3metric}
ds^2=\bar{g}_{\mu\nu}dx^\mu dx^\nu=\ell^2(-
\cosh^2{\rho}d\tau^2+\sinh^2{\rho}d\phi^2+d\rho^2)
\end{equation}
where the radius is related to $\Lambda$ by
\begin{equation}
\ell^{-2}=-\Lambda
\end{equation}
The Riemann tensor, Ricci tensor and Ricci scalar of the $AdS_3$ are:
\begin{equation}
\bar{R}_{\mu\alpha\nu\beta} = \Lambda (\bar{g}_{\mu\nu}
\bar{g}_{\alpha\beta} - \bar{g}_{\mu\beta} \bar{g}_{\alpha\nu})  ,
\quad \bar{R}_{\mu\nu} = 2 \Lambda \bar{g}_{\mu\nu}  , \quad \bar{R}
= 6 \Lambda  ,
\end{equation}
The metric (\ref{AdS3metric}) has isometry group $SL(2,\mathbb{R})_L\times SL(2,\mathbb{R})_R$. The $SL(2,\mathbb{R})_L$ generators
are
\begin{eqnarray}L_0 &= &  i \partial_u ~, \\ L_{-1} &= &  i e^{-iu} \left[ {
\cosh 2 \rho \over \sinh 2 \rho } \partial_u - { 1 \over \sinh 2
\rho}
\partial_v +  { i \over 2} \partial_\rho \right] ~, \\ L_{1} &= &  i
e^{iu} \left[ { \cosh 2 \rho \over \sinh 2 \rho } \partial_u - { 1
\over \sinh 2 \rho} \partial_v -  { i \over 2} \partial_\rho
\right]\end{eqnarray} where $u \equiv \tau + \phi, v \equiv
\tau-\phi$. The $SL(2,\mathbb{R})_R$ generators $\{\bar{L}_0,
\bar{L}_{\pm1}\}$ are given by the above expressions with $u
\leftrightarrow v$. The normalization of the $SL(2,\mathbb{R})$
algebra is
\begin{equation}[L_0,
L_{\pm1}] = \mp L_{\pm1} , ~~[L_1, L_{-1}] = 2L_0 .\end{equation}
The quadratic Casimir of $SL(2,\mathbb{R})_L$ is $L^2 = {1\over 2}
(L_1L_{-1} + L_{-1}L_1)- L^2_0$. When acting on scalars,
$L^2+\bar{L}^2=-{\ell^2\over2}\bar{\nabla}^2$ \cite{juandy}.

\subsection{Chiral gravity}
As shown by Brown and Henneaux \cite{brown}, quantum gravity on
asymptotically $AdS_3$ spacetimes with appropriate boundary
conditions is described by a 2D CFT which lives on the boundary.
They computed the total central charge of the CFT and found
$c_L+c_R=3\ell/G$. Simplified calculations were given in
\cite{Henningson:1998gx}\cite{de Haro:2000xn}\cite{balas}. The
difference $c_L-c_R$ corresponds to the diffeomorphism anomaly. In
reference \cite{Kraus:2005zm}\cite{kraus} it is shown that
 $c_L-c_R=-\frac{3}{\mu G}.$
 In summary, we have
\begin{equation}\label{cl}
c_L=\frac{3\ell}{2G}(1-\frac{1}{\mu \ell}) \qquad
c_R=\frac{3\ell}{2G}(1+\frac{1}{\mu \ell}) \end{equation} In order
that both  central charges are non-negative, we must have, as was
also noticed in \cite{Solodukhin:2005ah},
\begin{eqnarray}\label{Central charge condition} \quad \mu \ell \geqslant
1
\end{eqnarray}
We note that had we chosen the opposite sign in front of the Einstein-Hilbert action the central charge would be negative. An interesting special case is
\begin{equation}\mu\ell=1\end{equation}
which implies
\begin{equation}\label{chiral}
c_L=0,\qquad c_R=\frac{3\ell}{G}. \end{equation} We will refer to
this theory as chiral gravity. If the chiral gravity is unitary it
can have only right-moving excitations.

\section{Gravitons in $AdS_3$}

In this section we describe the linearized excitations around background $AdS_3$ metric $\bar{g}_{\mu\nu}$. Expanding
\begin{equation}
g_{\mu\nu} = \bar{g}_{\mu\nu} + h_{\mu\nu}
\end{equation}
with $h_{\mu\nu}$ small, the linearized
 Ricci tensor and Ricci scalar are \cite{DeserTekin2}
 \cite{Olmez:2005by},
\begin{eqnarray}
R_{\mu\nu}^{(1)}& =& \frac{1}{2} (- \bar{\nabla}^2  {h}_{\mu\nu} -
\bar{\nabla}_{\mu}  \bar{\nabla}_{\nu}  h + \bar{\nabla}^{\sigma}
\bar{\nabla}_{\nu}
 h_{\sigma\mu} + \bar{\nabla}^{\sigma}  \bar{\nabla}_{\mu}  h_{\sigma\nu})\\
R^{(1)} &\equiv& (R_{\mu\nu}  g^{\mu\nu})^{(1)} = - \bar{\nabla}^2 h
+ \bar{\nabla}_{\mu} \bar{\nabla}_{\nu}  h^{\mu\nu} - 2 \Lambda h.
\label{calgtr}\end{eqnarray} The leading terms in $\cal G$ and $C$
are
\begin{eqnarray}
\mathcal{G}_{\mu\nu}^{(1)} & =& R_{\mu\nu}^{(1)} - \frac{1}{2}
\bar{g}_{\mu\nu}
R^{(1)} - 2 \Lambda  {h}_{\mu\nu}  , \\
C^{(1)}_{\mu\nu } & = & \epsilon_\mu^{\,\,\alpha\beta}
{\bar{\nabla}}_{\alpha}  \left( R_{\beta\nu}^{(1)} - \frac{1}{4}  \bar{g}_{\beta\nu}  R^{(1)} - 2  \Lambda
h_{\beta\nu} \right)  .
\end{eqnarray}
We note $\overline{\textrm{Tr}} C^{(1)}=0 $ and the Bianchi identity implies
 $\bar{\nabla}^\mu{\cal G}^{(1)}_{\mu\nu}=\bar{\nabla}^\mu C^{(1)}_{\mu\nu}=0\
 $.
The linearized equations of motion are then
\begin{equation}
{\cal G}^{(1)}_{\mu\nu}+\frac{1}{\mu}C^{(1)}_{\mu\nu}=0
\end{equation}
Tracing this equation yields $\overline{\textrm {Tr}} {\cal
G}^{(1)}=-\frac{1}{2}R^{(1)}=0$. So the equation of motion becomes
\begin{equation}\label{calgfirst}
{\cal G}_{\mu\nu}^{(1)}+\frac{1}{\mu}\epsilon_\mu\,
^{\alpha\beta}\bar{\nabla}_\alpha{\cal G}_{\beta\nu}^{(1)}=0
\end{equation}
where in terms of $h_{\mu\nu}$
\begin{equation}\label{calgandh}
{\cal G}_{\mu\nu}^{(1)}=\frac{1}{2} (- {\bar{\nabla}^2} {h}_{\mu\nu}
- {\bar{\nabla}}_{\mu}  {\bar{\nabla}}_{\nu}  h +
{\bar{\nabla}}^{\sigma}  {\bar{\nabla}}_{\nu}  h_{\sigma\mu} +
{\bar{\nabla}}^{\sigma}  {\bar{\nabla}}_{\mu}  h_{\sigma\nu})
-2\Lambda h_{\mu\nu}
\end{equation}

Now we fix the gauge. We define $\tilde{h}_{\mu\nu} \equiv
h_{\mu\nu}-\bar{g}_{\mu\nu}h$, which gives $\tilde{h}=-2h$. Plugging
$h_{\mu\nu}=\tilde{h}_{\mu\nu}-\frac{1}{2}\bar{g}_{\mu\nu}\tilde{h}$
into (\ref{calgtr}) and setting it to zero gives
\begin{equation}
\bar{\nabla}_{\mu}\bar{\nabla}_\nu\tilde{h}^{\mu\nu}=-\Lambda
\tilde{h}
\end{equation}
Thus, the gauge
\begin{equation}
\bar{\nabla}_\mu\tilde{h}^{\mu\nu}=0
\end{equation} together with the linearized equation of motion
implies tracelessness of $h_{\mu\nu}$: $\tilde{h}=-2h=0$. This gauge
is equivalent to the harmonic plus traceless gauge
$\bar{\nabla}_{\mu}h^{\mu\nu}=h=0$. Noting that
\begin{equation}\label{cmt}
[\bar{\nabla}_\sigma,\bar{\nabla}_\mu]h^\sigma_{\nu}=\bar{R}^\sigma\,_{\lambda\sigma\mu}h^{\lambda}_\nu-\bar{R}^\lambda\,_{\nu\sigma\mu}h^\sigma_\lambda=3\Lambda
h_{\mu\nu}-\Lambda h g_{\mu\nu}\end{equation} and imposing the gauge
condition, (\ref{calgandh}) is just
\begin{equation}\label{emh}
{\cal G}_{\mu\nu}^{(1)}={1\over 2}
(-{\bar{\nabla}^2}h_{\mu\nu}+2\Lambda h_{\mu\nu}\ )\ .
\end{equation}
The equation of motion (\ref{calgfirst}) thus becomes
\begin{equation}\label{eomh}
(\bar{\nabla}^2+\frac{2}{\ell^2})(h_{\mu\nu}+\frac{1}{\mu}\epsilon_\mu\,
^{\alpha\beta}\bar{\nabla}_\alpha h_{\beta\nu})=0
\end{equation}

Define three mutually commuting operators $(\mathcal{D}^{L},\mathcal{D}^{R},\mathcal{D}^{M})$:
\begin{eqnarray}(\mathcal
D^{L/R})_\mu\,^\beta \equiv
\delta_\mu\,^\beta\pm\ell\epsilon_\mu\,^{\alpha\beta}\bar{\nabla}_\alpha,\qquad \textrm{and}\qquad (\mathcal{ D}^{M})_\mu\,^\beta\equiv
\delta_\mu\,^\beta+\frac{1}{\mu}\epsilon_\mu\,^{\alpha\beta}\bar{\nabla}_\alpha
\end{eqnarray}
where the meaning of superscripts will become clear presently.
The equations of motion (\ref{calgfirst}) can then be written as
\begin{equation}\label{eomfactor}
(\mathcal{D}^{L}\mathcal{D}^{R}\mathcal{D}^{M}h)_{\mu\nu}=0
\end{equation}
where we have used the linearized Bianchi identity. Since the three operators commute
equation (\ref{eomfactor}) has three branches of solutions.
First, the massive gravitons $h^{M}_{\mu\nu}$ given by
\begin{equation}\label{eommassiv}
(\mathcal{D}^{M}h^{M})_{\mu\nu}=h^{M}_{\mu\nu}+\frac{1}{\mu}\epsilon_\mu\,
^{\alpha\beta}\bar{\nabla}_\alpha h^{M}_{\beta\nu}=0
\end{equation}
are solutions special for TMG. The other two branches are massless gravitons which are also solutions of Einstein gravity: ${\cal G}_{\mu\nu}^{(1)}=0$. The left-mover $h^{L}_{\mu\nu}$ and right-mover $h^{R}_{\mu\nu}$ have different first order equations of motion:
\begin{equation}\label{eomleft}
(\mathcal{D}^{L}h^{L})_{\mu\nu}=h^{L}_{\mu\nu}+\ell\epsilon_\mu\,
^{\alpha\beta}\bar{\nabla}_\alpha h^{L}_{\beta\nu}=0 \qquad (\mathcal{D}^{R}h^{R})_{\mu\nu}=h^{R}_{\mu\nu}-\ell\epsilon_\mu\,
^{\alpha\beta}\bar{\nabla}_\alpha h^{R}_{\beta\nu}=0
\end{equation}
Note that the components of (\ref{eommassiv}) and (\ref{eomleft}) tangent to the
$AdS_3$ boundary relate those components of $h_{\mu\nu}$ to their falloff at infinity. This could be used to directly infer their conformal weights but we will instead just find the full solutions.

Next we solve for the three branches of solutions. Define linear
operator $(\tilde{\mathcal{D}}^{M})_\mu\,^\beta\equiv
\delta_\mu\,^\beta-\frac{1}{\mu}\epsilon_\mu\,^{\alpha\beta}\bar{\nabla}_\alpha$,
which commutes with $\mathcal{D}^{M}$ defined earlier. Applying
$\tilde{\cal D}^M$  on ({\ref{calgfirst}}), we get a second order
equation,
\begin{equation}\label{calgsec}
(\tilde{\mathcal{D}}^M\mathcal{D}^M
\mathcal{G}^{(1)})_{\mu\nu}=(\mathcal{D}^M\tilde{\mathcal{D}}^M\mathcal{G}^{(1)})_{\mu\nu}=-\frac{1}{\mu^2}[\bar{\nabla}^2-(\mu^2+3\Lambda)]\mathcal{G}^{(1)}_{\mu\nu}=0
\end{equation} where we have used the linearized Bianchi identity.
Note that $\tilde{\cal D}^M{\cal G}=0$ is
just the linearized gravitational wave equation if we exchange $\mu$
for $-\mu$. So solutions of TMG with both signs of $\mu$ are
 solutions  of (\ref{calgsec}). It can conversely be shown that all solutions
 of (\ref{calgsec}) are solutions of TMG for one sign of $\mu$ or the other.

Rewrite the Laplacian acting on rank two tensors in terms of the sum
of two $SL(2,\mathbb{R})$ Casimirs:
\begin{equation}\bar{\nabla}^2h_{\mu\nu}=-[{2\over\ell^ 2}(L^2+\bar{L}^2)+{6\over\ell^ 2}]h_{\mu\nu}\end{equation}
 ${\cal G}^{(1)}_{\mu\nu}$ can be written as
\begin{equation}
{\cal G}^{(1)}_{\mu\nu}=[{1\over
\ell^ 2}(L^2+\bar{L}^2)+\frac{2}{\ell^ 2}]h_{\mu\nu}\ .\end{equation}
Thus (\ref{calgsec}) becomes
\begin{equation}\label{hhbarsec}
[-\frac{2}{\ell^ 2}(L^2+\bar{L}^2)-\frac{3}{\ell^ 2}-\mu^2][{1\over
\ell^ 2}(L^2+\bar{L}^2)+\frac{2}{\ell^ 2}]h_{\mu\nu}=0\ .
\end{equation}
This allows us to use the $SL(2,\mathbb{R})_L\times SL(2,\mathbb{R})_R$ algebra to classify the
solutions of (\ref{calgsec}). Consider states with weight
$(h,\bar{h})$:
\begin{equation}L_0|\psi_{\mu\nu} \rangle=
h|\psi_{\mu\nu}\rangle ,\qquad
\bar{L}_0|\psi_{\mu\nu}\rangle=\bar{h}|\psi_{\mu\nu}\rangle.
\end{equation}
From the explicit form of the generators, we see
\begin{equation}\label{solpsi}
\psi_{\mu\nu}=e^{-ihu-i\bar{h}v}F_{\mu\nu}(\rho)
\end{equation}
Now let's specialize to primary states $|\psi_{\mu\nu} \rangle$
which obey $L_1|\psi_{\mu\nu}
\rangle=\bar{L}_1|\psi_{\mu\nu}\rangle=0.$ These conditions plus the
gauge conditions give $h-\bar{h}=\pm 2$  and
\begin{eqnarray}\label{solfuv}F_{\mu\nu}(\rho)=f(\rho)\left(\begin{array}{ccc}
                                         1 & {h-\bar{h}\over2}& {i\over\sinh(\rho)\cosh(\rho)} \\
                                          {h-\bar{h}\over2} & 1 & {i(h-\bar{h})\over2\sinh(\rho)\cosh(\rho)} \\
                                         {i\over\sinh(\rho)\cosh(\rho)} &{i(h-\bar{h})\over2\sinh(\rho)\cosh(\rho)}   & - {1\over\sinh^2(\rho)\cosh^2(\rho)} \\
                                       \end{array}\right)\end{eqnarray}
                                       where
\begin{equation}\partial_\rho
f(\rho)+\frac{(h+\bar{h})\sinh^2{\rho}-2\cosh^2{\rho}}{\sinh{\rho}\cosh{\rho}}f(\rho)=0\
\end{equation}  for the primary states. The solution is
\begin{equation}\label{solf}
f(\rho)=(\cosh{\rho})^{-(h+\bar{h})}\sinh^2{\rho}
\end{equation}We see that
$\sqrt{-\bar{g}}\psi^{\mu\nu*}\psi_{\mu\nu}={4\sinh\rho}~{(\cosh\rho)^{-3-2(h+\bar{h})}}.$
Also using $L^2|\psi_{\mu\nu}\rangle=-h(h-1)|\psi_{\mu\nu}\rangle$
for primaries, the primary weights $(h,\bar{h})$ obey:
\begin{equation}[2h(h-1)+2\bar{h}(\bar{h}-1)-3-\mu^2\ell^ 2][h(h-1)+\bar{h}(\bar{h}-1)-2]=0, \quad h-\bar{h}=\pm 2\end{equation}
There are two branches of solutions. The first branch has
$h(h-1)+\bar{h}(\bar{h}-1)-2=0$, which gives:
\begin{equation}
h={3\pm1\over2},\,\bar{h}={-1\pm1\over2}\quad \hbox{or}\quad h={-1\pm1\over2},\,\bar{h}={3\pm1\over2}
\end{equation}
These are the solutions that already appear in Einstein gravity. The
solutions with the lower sign will blow up at the infinity, so we
will only keep the upper ones corresponding to weights $(2,0)$ and
$(0,2)$. We will refer to these as left and right-moving massless
gravitons.

The second branch has $2h(h-1)+2\bar{h}(\bar{h}-1)-3-\mu^2\ell^2=0$,
which gives:
\begin{eqnarray}\label{mup}
h=\frac{3}{2}\pm\frac{\mu \ell }{2} &\qquad&
\bar{h}=-\frac{1}{2}\pm\frac{\mu \ell}{2}\\
\label{mun}\textrm{or}\qquad h=-\frac{1}{2}\pm\frac{\mu
\ell}{2}&\qquad&\bar{h}=\frac{3}{2}\pm\frac{\mu \ell }{2}
\end{eqnarray}
where (\ref{mup}) are the solutions of (\ref{calgfirst}), and
(\ref{mun}) are the solutions of (\ref{calgfirst}) with $\mu$
replaced by $-\mu$. In the case of interest $\mu\ge1/\ell,$ only the
solutions with the plus signs in (\ref{mup}) will not blow up at the
infinity. Hence the relevant solutions corresponding to massive
gravitons are
\begin{equation}\label{ffd}h=\frac{3}{2}+\frac{\mu \ell}{2},~~~~~~
\bar{h}=-\frac{1}{2}+\frac{\mu \ell}{2}\ .\end{equation} Descendants
are obtained by simply applying $L_{-1}$ and $\bar{L}_{-1}$ on the
primary $| \psi_{\mu\nu}\rangle\ .$

 Note that for chiral gravity at $\mu \ell =1,$ with $c_L=0,~c_R={3l\over G}$, (\ref{ffd}) becomes
 $h=2,\,\bar{h}=0$. Furthermore the wave function for the massive graviton becomes identical to that of the left-moving massless graviton.
In fact, we can eliminate the $(2,0)$ modes with the residual
gauge transformation
 \begin{eqnarray}\epsilon_t&=&e^{-2iu}\frac{i\sinh^4(\rho)}{6\cosh^2(\rho)}~,\\
 \epsilon_\phi&=&e^{-2iu}{-i\sinh^2(\rho)\left(2+\cosh^2(\rho)\right)\over6\cosh^2(\rho)}~,\\
 \epsilon_\rho&=&e^{-2iu}{\sinh(\rho)\left(1+2\cosh^2(\rho)\right)\over6\cosh^3(\rho)}~,\end{eqnarray}
which satisfies\begin{eqnarray}  \bar{\nabla}_\mu
\epsilon_\nu+\bar{\nabla}_\nu
\epsilon_\mu+\psi^{(2,0)}_{\mu\nu}&=&0\ .\end{eqnarray} Note that
this gauge transformation does not vanish at the boundary, so
whether or not the $(2,0)$ solution should be regarded as gauge
equivalent to the vacuum depends on the so-far-unspecified boundary
conditions.

 \section{Positivity of energy}

\subsection{BTZ black holes}

The addition of the Chern-Simons term to the bulk action requires
additional surface terms which in turn modify the definition of
energy \cite{Deser:1982vy}. These surface terms are non-vanishing in
general even for those solutions which satisfy the Einstein equation
--- namely BTZ black holes and massless gravitons.
 For Einstein metrics, the mass $M(\mu)$ and
angular momentum $J(\mu)$ at general coupling $\mu$ are related to
their values at $\mu=\infty$ by \cite{Moussa:2003fc}
\cite{Kraus:2005zm},
\begin{equation}\label{kko}\ell M(\mu)=\ell M(\infty)+{J(\infty) \over \mu \ell } \end{equation}
\begin{equation} J(\mu)=J(\infty)+{M(\infty) \over \mu  } \end{equation}

The bound $\ell M(\infty) \ge |J(\infty)|$ then  implies positivity
of energy for
 Einstein metrics when $\mu \ell >1$. When $\mu\ell<1$, the energy of the black holes can be negative
 (There is some discussion in this region in
 \cite{Park:2006gt}\cite{Park:2006fp}, where signs of instability also appear).
  Note that  for chiral gravity at $\mu \ell=1$ we have
\begin{equation}\ell M({1 \over \ell} )=J({1 \over \ell}) .\end{equation}
This can be interpreted as the statement that all Einstein
geometries are right-moving.

Now let's compute the entropy of the black hole, assuming the
existence of a unitary dual CFT, following \cite{krauslarsen}
\cite{Kraus:2005zm}. The inner and outer horizons are at
\begin{equation}r_\pm=\sqrt{{2G\ell(\ell M(\infty)+J(\infty))}}\pm\sqrt{{2G\ell(\ell M(\infty)-J(\infty))}}\end{equation}
and do not depend on $\mu$. In terms of these the macroscopic
formula for the entropy, including a contribution from the
Chern-Simons term,  is \cite{Solodukhin:2005ah} \cite{Sahoo:2006vz}
\cite{Park:2006gt} \cite{Tachikawa:2006sz}.
\begin{equation}\label{sbh}S_{BH}(\mu)={\pi \over 2G}(r_++{1 \over \mu\ell}r_-)\end{equation}
The left and right temperatures of the black hole are determined by periodicities and also
do not depend on $\mu$. They are
\begin{equation}T_L={r_+-r_-\over 2\pi \ell^2},~~~T_R={r_++r_- \over 2 \pi \ell^2}\end{equation}
In terms of these quantities, the microscopic Cardy formula for the entropy is
\begin{equation}S_{BH}(\mu)={\pi^2c_L(\mu)T_L \ell\over 3}+ {\pi^2c_R(\mu)T_R\ell\over 3} \end{equation}
Using formula (\ref{cl}) for the central charges one readily finds that this agrees with the macroscopic result (\ref{sbh}).

\subsection{Gravitons}
For $\mu \ell > 1$  the weights $(h,\bar h)$ of the massive
gravitons are positive. The energy of the massive gravitons is
proportional to these weights but with a possible minus sign. In
particular if the overall sign of the action is changed, so is the
energy, but the equations of motion and hence the weights are
unaffected. In \cite{Deser:1981wh} \cite{Deser:1982vy}
\cite{Deser:1982sw}, it was shown that with the sign taken herein
but no cosmological constant massive gravitons have negative energy.
In this section we redo this analysis for the case of negative
cosmological constant, by constructing the Hamiltonian
\cite{Buchbinder:1992pe}.

 The fluctuation $h_{\mu\nu}$ can be decomposed as
 \begin{eqnarray}h_{\mu\nu}&=&h^{M}_{\mu\nu}+h^L_{\mu\nu}+h^R_{\mu\nu}\end{eqnarray} where we use the subscript $M$ to represent the $({3+\mu \ell \over2},{-1+\mu \ell \over2})$ primary and their descendants,
 $L$ to represent the $(2,0)$ primary and their descendants, and $R$ to represent the $(0,2)$ primary and their
 descendants. We will call them massive modes, left-moving modes and
 right-moving modes hereafter.

Up to total derivatives, the quadratic action of $h_{\mu\nu}$ is
\begin{eqnarray}
S_2&=&-\frac{1}{32\pi G}\int d^3x \sqrt{-g} h^{\mu\nu}({\cal G}^{(1)}_{\mu\nu}+\frac{1}{\mu}C^{(1)}_{\mu\nu})\\
&=&\frac{1}{64\pi G}\int d^3x
 \sqrt{-g}\{-\bar{\nabla}^\lambda h^{\mu\nu}\bar{\nabla}_\lambda
 h_{\mu\nu}+\frac{2}{
\ell^ 2}h^{\mu\nu}h_{\mu\nu}-\frac{1}{\mu}\bar{\nabla}_\alpha
h^{\mu\nu}\epsilon_\mu\,^{\alpha\beta}(\bar{\nabla}^2+\frac{2}
{\ell^ 2})h_{\beta\nu}\}\nonumber\end{eqnarray} The momentum conjugate to
$h_{\mu\nu}$ is
\begin{equation}
\Pi^{(1)\mu\nu}=-{\sqrt{-g}\over64\pi G}
\left(\bar{\nabla}^0(2h^{\mu\nu}+{1\over\mu}\epsilon^{\mu\alpha}\,_{\beta}\bar{\nabla}_\alpha
h^{\beta \nu}) -{1\over\mu}\epsilon_\beta\,^{0\mu
}(\bar{\nabla}^2+{2\over \ell^ 2})h^{\beta\nu}\right)\end{equation}
Using the equations of motion we find, \begin{eqnarray}
\Pi_M^{(1)\mu\nu}&=&{\sqrt{-g}\over64\pi
G}(-\bar{\nabla}^0h^{\mu\nu}+{1\over\mu}(\mu^2-{1\over
\ell^ 2})\epsilon_\beta\,^{0\mu}h_M^{\beta\nu})\\
\Pi_L^{(1)\mu\nu}&=&-{\sqrt{-g}\over64\pi G}(2-{1\over\mu
\ell}) \bar{\nabla}^0h_L^{\mu\nu}\\
\Pi_R^{(1)\mu\nu}&=&-{\sqrt{-g}\over64\pi G}(2+{1\over\mu
\ell})\bar{\nabla}^0h_R^{\mu\nu}\end{eqnarray}

Because we have up to three time derivatives in the Lagrangian,
using the Ostrogradsky method we should also introduce
$K_{\mu\nu}\equiv\bar{\nabla}_0h_{\mu\nu}$ as a canonical variable,
whose conjugate momentum is
\begin{equation}\Pi^{(2)\mu\nu}=\frac{-\sqrt{-g}g^{00}}{64\pi G\mu}\epsilon_\beta\,^{\lambda\mu}\bar{\nabla}_\lambda h^{\beta\nu}\end{equation}
again using equations of motion we have
\begin{eqnarray}\Pi^{(2)\mu\nu}_M&=&\frac{-\sqrt{-g}g^{00}}{64\pi G}h_M^{\mu\nu}\\
\Pi^{(2)\mu\nu}_R&=&\frac{-\sqrt{-g}g^{00}}{64\pi G\mu
\ell}h_L^{\mu\nu}\\\Pi^{(2)\mu\nu}_L&=&\frac{\sqrt{-g}g^{00}}{64\pi
G\mu \ell}h_R^{\mu\nu}\end{eqnarray}
The Hamiltonian is then
\begin{equation}H=\int d^3x\bigl(\dot{h}_{\mu\nu}\Pi^{(1)\mu\nu}+\dot{K}_{i\mu}\Pi^{(2)i\mu}-S\bigr)\end{equation}
Specializing to linearized gravitons, and using their equations of
motion,  we then have the energies
\begin{eqnarray}\label{energy}E_M&=&\frac{1}{\mu}(\mu^2-{1\over
\ell^ 2})\int d^3x{\sqrt{-g}\over64\pi G}\epsilon_\beta\,^{0\mu}h_M^{\beta\nu}\dot{h}_{M\mu\nu}\\
E_L&=&(-1+{1\over\mu
\ell})\int d^3x{\sqrt{-g}\over32\pi G}\bar{\nabla}^0h_L^{\mu\nu}\dot{h}_{L\mu\nu}\\
E_R&=&(-1-{1\over\mu \ell})\int d^3x{\sqrt{-g}\over32\pi
G}\bar{\nabla}^0h_R^{\mu\nu}\dot{h}_{R\mu\nu}\end{eqnarray}

All the integrals above are negative, as can be shown by plugging in the solutions (\ref{solpsi}), (\ref{solfuv}) and (\ref{solf}) for primaries and by using the $SL(2,\mathbb{R})$ algebra for descendants.
 For the massive mode, the energy is negative when $\mu
\ell
>1$ (it becomes infinitely negative at $\mu=\infty$) and positive when
$\mu \ell <1$. The energy of left-moving modes is positive when $\mu
\ell
>1$, and negative when $\mu \ell <1$. The energy of right-moving modes is always
positive. $\mu \ell =1$ is a critical point, where the energy of
both massive modes and the left-moving modes becomes zero. Note
$E_L$ and $E_R$ are consistent with (\ref{kko}).

Recall that positivity of central charges requires that $\mu\ell \geq 1$, so the only possibility for avoiding negative energy is to take the chiral gravity theory with $\mu \ell =1$. In that case, we see that
massive and left-moving gravitons carry no energy, and might perhaps be regarded as pure gauge.

\section{Conclusion and Discussion}

In this note, we investigated the cosmological TMG, and found that
the theory can be at most sensible at $\mu\ell=1$. At this special
point, the theory is completely chiral. In order to show that the
chiral theory is classically sensible asymptotic boundary conditions
which consistently eliminate the infinite degeneracy of zero-energy
excitations must be specified. One must also prove a non-linear
positive energy theorem. Renormalizability must be addressed to
define the quantum theory.

At the classical level the chiral structure might enable an exact
solution of the theory. Some exact non-Einstein solutions of TMG with arbitrary
$\mu$ have been found in
\cite{Percacci:1986ja}-\cite{Bouchareb:2007yx}. Should there turn
out to also be a consistent quantum theory it would be interesting
to find the chiral CFT dual. Towards this end the approach of
\cite{Witten:2007kt} may prove useful.

\section*{Acknowledgements}
We thank E. Witten and X. Yin for helpful discussions. The work is supported by
DOE grant DE-FG02-91ER40654. WS is also supported by CSC. WS would
like to thank the hospitality of the physics department of Harvard
University.
\appendix{\bf APPENDIX: Energy-momentum pseudotensor}

In this appendix we show that the energy defined from the energy
momentum pseudotensor for massive gravitons can be negative. For
simplicity we specialize to $\Lambda=0$: for $\mu \ell \gg 1$ the
Compton wavelength of the gravitons is much shorter than the $AdS_3$
radius so the latter can be locally ignored.

Let us first review how the energy-momentum pseudo tensor at quadratic order is defined
without the Chern-Simons term or cosmological constant term. The
full Einstein equation in the presence of matter is
\begin{equation}G_{\mu\nu}=16\pi G T^M_{\mu\nu}\end{equation}
Expanding the metric $g_{\mu\nu}=\eta_{\mu\nu}+h_{\mu\nu}$, we have
through quadratic order
\begin{equation}G^{(1)}_{\mu\nu}=-G^{(2)}_{\mu\nu}+16\pi G
T^M_{\mu\nu}\,\end{equation} where $G^{(1)}_{\mu\nu}$ and
$G^{(2)}_{\mu\nu}$ are terms linear and quadratic in $h_{\mu\nu}$.
The energy momentum pseudo tensor defined as
\begin{equation}t_{\mu\nu}=-{1\over 16\pi
G}G^{(2)}_{\mu\nu},\quad\hbox{with}\quad
\partial^\mu t_{\mu\nu}=0,
\end{equation}
sources the Newtonian part of the gravitational potential in the
same way that the matter stress tensor does. When adding the
Chern-Simons term, the energy momentum pseudo tensor is similarly
defined as
\begin{equation}t_{\mu\nu}=-{1\over 16\pi G}(G^{(2)}_{\mu\nu}+{1\over \mu}C^{(2)}_{\mu\nu}).
\end{equation}
In flat background, the linearized equation of motion
(\ref{calgfirst}) becomes
\begin{equation}\label{eqmflat}\partial^2h_{\mu\nu}+{1\over\mu}\epsilon_\mu\,^{\alpha\beta}\partial_\alpha\partial^2 h_{\beta\nu}=0\end{equation}
under the harmonic plus traceless gauge. For a plane wave solution
in the form of $h_{\mu\nu}(k)={1\over\sqrt{2 k_0}}e^{-ik\cdot
x}e_{\mu\nu}(k)$, the gauge conditions and the equations of motion
are
\begin{eqnarray}k^\mu e_{\mu\nu}(k)&=&0,\,e^\mu\,_\mu(k)=0
\\ k^2[e_{\mu\nu}(k)&-&{ik_\alpha\over\mu}\epsilon_\mu\,^{\alpha\beta}e_{\beta\nu}(k)]=0.\end{eqnarray}
The $k^2=0$ solution is pure gauge. So
$e_{\mu\nu}(k)-{ik_\alpha\over
\mu}\epsilon_\mu\,^{\alpha\beta}e_{\beta\nu}(k)=0$, which implies
$(k^2+\mu^2)e_{\mu\nu}(k)=0$. When $k_\mu=(\mu,0,0)$, the positive
energy solution is
\begin{equation}\label{solutionflat} h_{\mu\nu}=\frac{e^{-i\mu \tau}}{\sqrt{2\mu}}\left(\begin{array}{ccc}
0 & 0 & 0 \\ 0 & 1 & i \\0 & i & -1
\\\end{array}\right)\end{equation}

It is convenient to define $e_\mu(\vec{0})=(0,1,i),$ such that
\begin{eqnarray}e_{\mu\nu}(\vec{0})&=&e_\mu(\vec{0})e_\nu(\vec{0}),\\ \hbox{and}\quad
e_{\mu\nu}(\vec{k})&=&e_\mu(\vec{k})e_\nu(\vec{k}),
\end{eqnarray}
 where $e_{\mu}(\vec{k})$ is obtained  from $e_\mu(\vec{0})$
by a boost.
The metric fluctuation can be expanded in Fourier modes,
\begin{equation}h_{\mu\nu}=\int d\vec{k}^2 {1\over\sqrt{2k_0}}\{ \alpha(\vec{k}) e_{\mu\nu}(\vec{k})e^{-ik\cdot x}+
 \alpha^\dag(\vec{k}) e^*_{\mu\nu}(\vec{k})e^{ik\cdot x}\}\end{equation}
 To calculate the energy momentum pseudo tensor, we will need the
 following,
\begin{eqnarray}t_{\mu\nu}&=&-{1\over 16\pi G}({\cal G}^{(2)}_{\mu\nu}+{1\over\mu}C^{(2)}_{\mu\nu})\\
{\cal
G}^{(2)}_{\mu\nu}&=&R^{(2)}_{\mu\nu}-{1\over2}\eta_{\mu\nu}R^{(2)}
\\R^{(2)}_{\mu\nu}&=& {1\over2} h^{\rho\sigma}\partial_\mu\partial_\nu h_{\rho\sigma}-
h^{\rho\sigma}\partial_\rho\partial_{(\mu}h_{\nu)\sigma}
+{1\over4}(\partial_\mu h_{\rho\sigma})\partial_\nu
h^{\rho\sigma}\\&+&(\partial^\sigma
h^\rho\,_\nu)\partial_{[\sigma}h_{\rho]\mu}+{1\over2}\partial_\sigma(h^{\rho\sigma}\partial_\rho
h_{\mu\nu})\\C^{(2)}_{\mu\nu}&=&\epsilon_{\mu}\,^{\alpha\beta}\partial_\alpha
(R^{(2)}_{\beta\nu}-{1\over4}\eta_{\beta\nu}R^{(2)})+h_{\mu\lambda}\epsilon^{\lambda\alpha\beta}\partial_\alpha
R^{(1)}_{\beta\nu}-\epsilon_\mu\,^{\alpha\beta}\Gamma^{\lambda(1)}_{\alpha\nu}R^{(1)}_{\beta\lambda}
\\ \Gamma^{\lambda(1)}_{\nu\alpha}&=&{1\over 2}\eta^{\lambda\rho}
(\partial_\alpha h_{\rho\nu}+\partial_\nu h_{\rho\alpha}-\partial_\rho h_{\alpha\nu})\\
R^{(1)}_{\mu\nu}&=&-{1\over2}\partial^2h_{\mu\nu}\end{eqnarray}

Using the equation of motion (\ref{eommassiv}), it simplifies to
\begin{eqnarray}G^{(2)}_{\mu\nu}&=&-{1\over4}(\partial_\mu h_{\rho\sigma})\partial_\nu
h^{\rho\sigma}-{\mu^2\over2}h^{\rho}\,_{\mu}
h_{\rho\nu}-{\mu^2\over8}\eta_{\mu\nu}h^{\rho\sigma}h_{\rho\sigma},\\C^{(2)}_{\mu\nu}&=&{\mu^2\over4}(3\mu
h_{\rho\mu}h^\rho\,_\nu+\epsilon_\mu\,^{\alpha\beta}h^\lambda\,_{\alpha,\nu}h_{\beta\lambda})\end{eqnarray}
up to total derivatives. So the energy is \begin{eqnarray}E&=&\int
d^2\vec{x}\,t_{00}
\\&=&\int
d^2\vec{k}{n(k)\over16\pi Gk_0}\{(k_0^2-{\mu^2\over2}-k_0\mu)
-{1\over2}\mu^2e_0(\vec{k})^*e_0(\vec{k})\}.\end{eqnarray} For
vanishing spatial momentum $n(\vec{k})\propto\delta^2(\vec{k})$, the
energy is just $-\mu$ times a positive normalization factor.


\end{document}